\documentstyle[12pt]{article}
\begin{document}

\begin{center}
{\Large {\bf Observed Asymmetry in {\boldmath 
        $\bar{p}p \to \pi^+K^-K^0\ /\ \pi^- K^+ \bar{K}^0$}} }\\[0.4cm]
{\Large {\bf and Relation to Reciprocity }}\\[2.4cm]
{\large P. K. Kabir}\footnote[1]{
e-mail address: pkk@virginia.edu}\\[0.4cm]
{\em J.W.Beams Physics Laboratory, 
     University of Virginia,
     Charlottesville, VA, USA }
\end{center}
\vskip1.4cm 
\centerline{\bf ABSTRACT} 
The  charge-asymmetry observed in   a   recent CPLEAR experiment   was
interpreted  by  the    authors    as   a   direct  observation     of
T-noninvariance. While   this    is the  simplest   and   most natural
inference, and the observed  effect agrees in sign and  magnitude  
with theoretical expectation, adherents of  T-invariance  may    argue 
that other interpretations are also possible. If $K^0$ and $\bar{K}^0$
are produced equally in $\bar{p}p$ annihilation, and T-invariance is
assumed to hold, the asymmetry observed in CPLEAR must be attributed to
TCP-noninvariance of kaon beta-decays. If  that were  the  case, the
charge-asymmetry in $K_S^0 \to \pi l\nu$ decays should be  three times
larger than the one observed for $K_L^0$ decays.

\newpage

\section{Introduction}

The CPLEAR collaboration has measured [1] a hitherto unreported C- and
CP-asymmetry in    $\bar{p}p$  annihilation  which,  under  reasonable
assumptions,  can be    identified with a    previously  predicted [2]
T-asymmetry.   Until now, there has been   no credible evidence of any
departure from reciprocity in  any reaction; also, questions have been
raised [3] about the  significance of the  test proposed in Ref.\ [2].
Therefore, it may be  useful  to critically examine the  circumstances
under   which  the  C-  and CP-asymmetry  reported   by  CPLEAR can be
interpreted  as a demonstration   of deviation from  T-invariance.  We
indicate further tests, and the conditions which must be satisfied for
the CP-asymmetry found by CPLEAR to be consistent with T-invariance.

\section{Expectation of T-Asymmetry}

The  departure  from CP-invariance  in   neutral kaon decays  has been
reliably established [4]   by a  number  of independent  measurements,
including a predicted  asymmetry [5] in $K_L  \to  \pi^+\pi^- e^+ e^-$
decays which has been  observed  recently [6].  However,  despite many
searches, there has been no clear  evidence of CP-noninvariance in any
phenomenon other than neutral kaon  decays. There,the observed effects
can be attributed entirely to $K^0-\bar{K}^0$ mixing, which could arise
from  CP-noninvariant interactions   much weaker  [7]  than   the weak
interactions responsible for the decay  of kaons. This may explain the
failure to see measureable CP-asymmetric effects in other phenomena.

Invariance  of physical  laws   under inversion [8]   of 4-dimensional
space-time --- which is not  {\em required} by Lorentz-invariance  but
obtains  in  most   Lorentz-invariant  theories with  further  minimal
analytic properties,    e.g.\  field  theories  described   by   local
Lagrangians   ---  can be given  a   consistent interpretation only if
space-time   inversion  PT  is accompanied   by  particle-antiparticle
conjugation C.  Within the  class of such TCP-invariant theories, lack
of  symmetry with respect to any  of the constituent operations, e.g.\
particle-antiparticle exchange C or ``combined inversion" CP, in which
space-coordinates          are     inverted   simultaneously      with
particle-antiparticle  interchange, {\em must}    be compensated by  a
corresponding  asymmetry  with respect  to one  or   more of the other
constituent operations, to preserve   the overall TCP- symmetry.    On
this basis,   Lee,  Oehme  and  Yang [9]   showed  that  the  possible
noninvariance with respect to space-inversion proposed [10] to explain
the ``tau-theta puzzle" necessarily required another presumed symmetry
to  be   broken;  they showed  that   observation    of the  suggested
P-noninvariant   effects   would  require  C-invariance   also  to  be
broken. An elegant way to preserve the symmetry of space, even if P is
abandoned,  suggested by  several authors  [11], is  to require  exact
CP-symmetry, in which  case TCP-invariance would  automatically assure
exact T-invariance  as well. The  subsequent discovery [4]  that CP is
not a valid symmetry in K-meson decays,  requires T-invariance also to
fail if TCP-invariance is to survive.  Following the discovery [12] of
parity- nonconservation,  searches [13] for T-noninvariance were based
largely on philosophical grounds: if physical laws are not indifferent
to space-inversion, perhaps they might  not be symmetric with  respect
to $t$-inversion either.  After the discovery of CP-nonconservation, the
search for  T-noninvariance  became a   logical  imperative.    Either
T-invariance would also fail, as TCP-invariance requires, or one would
face the even greater challenge of TCP-noninvariance.

As  long as deviations from  CP-symmetry are  confined to neutral kaon
decays and associated effects, the only place where one has a definite
expectation   of   seeing  T-noninvariance   must  be   in   the  same
phenomena. Furthermore, if TCP-invariance  is valid, the  observed CP-
noninvariance manifested   in   neutral kaon    decays {\em  must}  be
accompanied by  corresponding deviations from  T-invariance, which  is
more   precisely   described   as   symmetry     with    respect    to
motion-reversal. TCP-invariance requires
\begin{equation}
  \label{eq:1}
(\tilde{a}_T | S | \tilde{b}_T )\ =\ ( b | S | a) 
\end{equation} %Eq. (1)
where $\tilde{c}$ represents the CP-transform  of the channel $c$  and
$c_T$  represents its time-reverse,  viz.\ the  channel  $c$ with  all
particle momenta and spins reversed. The requirement of CP-invariance:
\begin{equation}
  \label{eq:2}
(\tilde{b} | S | \tilde{a} )\  =\  ( b | S | a )
\end{equation} %Eq. (2)
taken together with Eq.\ (1), would require that
\begin{equation}
  \label{eq:3}
(\tilde{a}_T | S | \tilde{b}_T )\ =\ 
(\tilde{b} | S | \tilde{a} )
\end{equation} %Eq. (3)
i.e.\ CP-invariance requires reciprocity  if TCP-invariance is  valid.
Conversely, if the requirement, Eq.\ (2), of CP-invariance {\em fails}
for a related pair of transition matrix-elements,  there {\em must} be
a corresponding failure of reciprocity in the same case [14].

We already mentioned   that a very  feeble CP-noninvariant interaction
contributing   to $K^0-\bar{K}^0$ mixing  suffices to  account for all
observed  CP-asymmetric   effects.   Therefore,  the  departure   from
T-invariance expected on the basis  of TCP-invariance must also appear
in $K^0-\bar{K}^0$ mixing. Departure from reciprocity would appear in 
a difference between  the  rates   of $\bar{K}^0\rightarrow  K^0$  and
$K^0\rightarrow\bar{K}^0$   transitions, expressed   by  a T-asymmetry
parameter[2,15]:
\begin{equation}
  \label{eq:4}
A_T\  =\ \frac{ P_{K\bar{K}} (\tau )\  -\   P_{\bar{K}K} (\tau ) }{
P_{K\bar{K}} (\tau )\  +\   P_{\bar{K}K} (\tau ) }
\end{equation} %Eq. (4)
which is found to be a  constant in  the generalized  Weisskopf-Wigner
approximation. Its value is given by
\begin{equation}
  \label{eq:5}
A_T^{\rm th}\ =\  2 {\rm Re} (\epsilon_S + \epsilon_L)\ =\
2 {\rm Re}  \langle K_L | K_S \rangle\, 
\end{equation} %Eq. (5)
to lowest order in the CP-nonconserving parameters  $\epsilon_{S,L}$,
defined by 
\begin{equation}
  \label{eq:6}
K_{S,L} \  \propto\ [1 + \epsilon_{S,L} ] K^0\ \pm
 [1 - \epsilon_{S,L} ] \bar{K}^0\, .
\end{equation} %Eq. 6
TCP-invariance requires [9] $\epsilon_S$ and $\epsilon_L$ to be equal;
on  that basis, the    value  of $A_T$  could   be  predicted  to   be
4Re$\epsilon  =  (6.4  \pm  1.2)\times  10^{-3}$ [17].    Even without
assuming any symmetry,  the last  quantity  on the right-hand side  of
Eq.\ (5) can be deduced by appeal to unitarity [18].   On the basis of
reasonable  assumptions, the most  relevant of which were subsequently
verified [17],   about upper limits   on minor modes  of  neutral kaon
decay, it  was shown  [2] that  the  expected T-asymmetry should  have
substantially the value predicted  for the TCP-invariant case, whether
that symmetry is assumed or not.

\section{CP-Asymmetry Measured by CPLEAR}

$\bar{p}p$   annihilations    into   [19]     $\pi^+K^-$``$K^0$"   and
$\pi^-K^+$``$\bar{K}^0$",   which are   expected   to  occur   equally
frequently by CP-invariance, were selected  by kinematic analysis, and
the frequencies of beta-decay  of the neutral  kaons were compared for
the two cases.  If  we accept the $\Delta  S=\Delta Q$ rule [20] which
requires  that $\pi^-e^+\nu$ and  $\pi^+e^-\bar{\nu}$  arise only from
$K^0$ and $\bar{K}^0$,  respectively, and  assume  that the two  decay
rates are equal, as   required  by TCP-invariance, then the   observed
$\pi^-e^+\nu$ and $\pi^+e^-\bar{\nu}$ rates at any time $\tau$ measure
the $K^0$  and $\bar{K}^0$ populations at  that  time.        Assuming
initial equality[22] of $K^0$ and $\bar{K}^0$ populations and survival
probabilities, any inequality between the observed annihilation rates
into:
\begin{equation}
  \label{eq:7}
p\bar{p} \to  \pi^+ K^- \{ \pi^+ e^-\bar{\nu} \}\  {\rm and}\ 
\pi^-K^+ \{ \pi^-e^+\nu \}
\end{equation} %Eq. (7)
must arise from a difference between $\bar{K}^0  \to K^0$ and $K^0 \to
\bar{K}^0$ transition rates. This is the conclusion drawn by CPLEAR.
  
The CP-asymmetry which they measure is:
\begin{equation}
  \label{eq:8}
A_l \ =\ 
\frac{ R [\pi^+ K^- \{\pi^+e^-\bar{\nu}\} ]\ -\ R [ \pi^-K^+ \{\pi^-e^+\nu\}] }
{R [\pi^+ K^-\{\pi^+e^-\bar{\nu}\} ]\ +\ R [ \pi^-K^+ \{\pi^-e^+\nu\} ]}
\end{equation} %Eq. (8)
In Eqs.\  (7)  and (8), the  $\pi  e\nu$ configurations in  braces are
observed as (delayed) end-products deduced kinematically to arise from
beta-decays of neutral kaons. Assuming the validity of the $\Delta S =
\Delta Q$ rule, this asymmetry can be written as:
\begin{equation}
  \label{eq:9}
A_l\ =\ \frac{ P_{K\bar{K}}(\tau ) R[ K^0\to \pi^-e^+\nu ]\ -\ 
P_{\bar{K}K}(\tau ) R[\bar{K}^0\to \pi^+ e^-\bar{\nu} ] }
{ P_{K\bar{K}}(\tau ) R[ K^0\to \pi^-e^+\nu ]\ +\ 
P_{\bar{K}K}(\tau ) R [\bar{K}^0\to \pi^+ e^-\bar{\nu} ] }\ .
\end{equation} %Eq. (9)
TCP-invariance requires that 
\begin{displaymath}
R [ K^0 \to \pi^-e^+\nu ]\  =\  R [ \bar{K}^0 \to \pi^+e^-\bar{\nu} ]\, ,
\end{displaymath} 
therefore, under the assumption of TCP-invariance, the CPLEAR asymmetry
becomes 
\begin{equation}
  \label{eq:10}
A_l^{\rm TCP}\ =\ \frac{ 
P_{K\bar{K}}(\tau )\ -\ P_{\bar{K}K}(\tau )}
{ P_{K\bar{K}}(\tau )\ +\ P_{\bar{K}K}(\tau ) }\ =\ A_T
\end{equation} %Eq. (10)
which is a measure of T-asymmetry  at the same  time as CP-asymmetry. 
Over a time-interval $\tau_S <  \tau < 20  \tau_S$, the observed 
asymmetry is consistent with being a constant, with a value reported 
as[1]
\begin{equation}
  \label{eq:11}
A_T^{\rm exp}\  =\     ( 6.6 \pm 1.3 ) \times 10^{-3}
\end{equation} %Eq. (11)
which agrees  with the theoretical prediction.   On the other hand, if
we insist on exact reciprocity,
\begin{equation}
  \label{eq:12}
P_{K\bar{K}}(\tau )\  =\ P_{\bar{K}K} (\tau )\, ,
\end{equation} %Eq. (12)
then  Eq.\ (9) reduces, for the case of exact T-invariance, to  
\begin{equation}
  \label{eq:13}
A_l^T\ =\ \frac{R[ K^0 \to \pi^-e^+\nu ]\ -\ R[ \bar{K}^0 \to \pi^+e^-
\bar{\nu} ]}{
R[ K^0 \to \pi^-e^+\nu ]\ +\ R[ \bar{K}^0 \to \pi^+e^- \bar{\nu} ] }
\end{equation} %Eq. (13)
and represents   a ( CP-   and )  CPT-violating effect.  The  observed
asymmetry  $A_l$, Eq.\ (8), requires the  beta-decay rate for $K^0 \to
\pi^-e^+\nu$ to exceed  that for $\bar{K}^0 \to \pi^+ e^-\bar{\nu}$ by
about 1.3$\%$, if exact T-invariance is imposed. If we parametrize the
deviation from TCP-invariance of kaon beta-decay amplitudes by setting
[23,24]:
\begin{equation}
  \label{eq:14}
\langle \pi^+ e^- \bar{\nu} | T | \bar{K}^0 \rangle\ =\
(1 + y ) \langle \pi^- e^+ \nu | T |K^0 \rangle
\end{equation} %Eq. (14)
where $y$  can be taken  to be real  without  loss of  generality, the
CP-asymmetry, Eq.\  (13), is given, to  lowest order  in $y$,  by -$y$; 
$y$ is therefore required to have the value:
\begin{equation}
  \label{eq:15}
y\  =\ -  ( 6.6 \pm 1.3 ) \times 10^{-3}
\end{equation} %Eq. (15)
if exact T-invariance is demanded. 

The  charge-asymmetry in $K_L^0 \to   \pi e\nu$ decays was  accurately
measured in several concordant experiments,  whose combined result  is
quoted as [17]:
\begin{equation}
  \label{eq:16}
\delta_l\ =\  ( 3.27 \pm 0.12 ) \times 10^{-3}
\end{equation} %Eq. (16) 
The phenomenological analysis  without assumption of any symmetry, but
assuming the validity of $\Delta Q = \Delta S$, yields [23]
\begin{equation}
  \label{eq:17}
\delta_{l,L}\   =\  2 {\rm Re} \epsilon_L - y\, .
\end{equation} %Eq. (17) 
The corresponding quantity for $K_S^0$ decays is
\begin{equation}
  \label{eq:18}
\delta_{l,S}\ =\  2 {\rm Re} \epsilon_S - y\, .
\end{equation} %Eq. (18)
T-invariance requires[25]   $\epsilon_S  = -  \epsilon_L$,  therefore
Eqs.\ (17) and (18) would constrain the leptonic charge-asymmetry from
$K_S^0$ decays to have the value:
\begin{equation}
\delta_{l,S}^T\ =\ - \delta_{l,L} - 2y\ =\  ( 9.9 \pm 1.3 )\times 10^{-3}\,,
\end{equation} %Eq. (19)
viz.\ {\em three} times the value, Eq.\ (16), for $K_L^0 \to \pi e\nu$
decays, if T-invariance  is to be  sustained. The CPLEAR data probably
contain the information required to confirm or refute this expectation
[26].  If  not, $\Phi$-decays    from  DA$\Phi$NE, which  provide    a
certified $K_S^0$  in   association with  each  $K_L^0$ decay,  should
provide a clean $K_S^0$ sample to test the unambiguous prediction (19)
required by the hypothesis of T-invariance.

\section{Conclusions}

The simplest interpretation of the CPLEAR  asymmetry, reported in Eq.\
(11), is that it  exhibits  the T-asymmetry predicted previously,  and
confirms  the sign and magnitude of  the expected effect. To this, the
logical objection may   be raised that   the CP-asymmetry measured  by
CPLEAR translates into the T-asymmetry   factor $A_T$ defined in  Eq.\
(4) {\em only} if the $\bar{p}p$ annihilation rates into $\pi^+K^-K^0$
and   $\pi^-K^+\bar{K}^0$ {\em and}  the   beta-decay   rates   for      
$K^0 \to \pi^-e^+ \nu$ and for $\bar{K}^0\to \pi^+e^-  \bar{\nu}$ are 
{\em assumed} to be equal. The latter is  required by TCP-invariance; 
but if one is  prepared to accept  TCP as an  article  of faith, then
T-noninvariance follows as
soon as   CP-invariance  fails   and  no  further    demonstration  is
required. Analysis  of  the CPLEAR asymmetry,  {\em  without} assuming
equality   of $K^0$  and $\bar{K}^0$   beta-decay  rates, shows  that,
subject   to    the  $\Delta Q   =  \Delta    S$   rule,  the leptonic
charge-asymmetry for $K_S^0 \to \pi e\nu$ decays should be three times
larger than the measured asymmetry for $K_L^0$ decays, if T-invariance
is valid.  Thus, it should not be too difficult to distinguish between
the simple interpretation  of the CPLEAR  charge-asymmetry as a direct
demonstration of T-noninvariance, and the desperate and radical resort
to TCP-noninvariance required to preserve T-invariance; these are the
only two alternatives unless one is willing to countenance  unequal
$production$ of $K^0$ and $\bar{K}^0$ in $\bar{p}p$ annihilations.
 
\section{Acknowledgments}

I thank G.V. Dass, G. Karl, and A. Pilaftsis for careful reading   of
the manuscript and for helpful suggestions to clarify its meaning.


\begin{thebibliography}{99}
\bibitem{20} There  are strong theoretical  arguments in favour of the
  $\Delta   S= \Delta Q$    rule  [21]. In   the  standard model,  the
  strangeness-changing  weak current  ($\bar{u}.s$)  necessarily obeys
  this rule. For analysis without this assumption, see Ref. [23].
\bibitem{24}     Non-vanishing  $y$    would  introduce  corresponding
  differences  between  $K^+ \to  \pi^0 e^+ \nu$   and  $K^- \to \pi^0
  e^-\nu$ decay amplitudes,  but available measurements [17]  have not
  attained  the precision    for  a significant  test.   Comparison of
  $K^{\pm}$ lifetimes with  a relative accuracy better  than $10^{-4}$
  could also provide useful limits on the value of $y$.
\bibitem{26} Formulas for the  complete time-distributions of leptonic
  decays,  including effects of   possible  $\Delta  Q = - \Delta   S$
  admixtures, are given in Ref.\ [23].
\end{thebibliography}
\end{document}